\documentstyle[prl,aps,psfig]{revtex}
\begin{document}
\draft
\twocolumn[\hsize\textwidth\columnwidth\hsize\csname @twocolumnfalse\endcsname
\title{Vortex stability of interacting Bose-Einstein condensates confined in
anisotropic harmonic traps}
\author{David L. Feder,$^{1,2}$ Charles W. Clark,$^2$ and Barry I.
Schneider$^3$}
\address{$^1$University of Oxford, Parks Road, Oxford OX1 3PU, U.K.}
\address{$^2$Electron and Optical Physics Division, National Institute of
Standards and Technology, Technology Administration, U.S. Department of
Commerce, Gaithersburg, MD, 20899}
\address{$^3$Physics Division, National Science Foundation, Arlington,
Virginia 22230}
\date{\today}
\maketitle
\begin{abstract}
Vortex states of weakly-interacting Bose-Einstein condensates confined in
three-dimensional rotating harmonic traps are investigated numerically at zero
temperature. The ground state in the rotating frame is obtained by propagating
the Gross-Pitaevskii (GP) equation for the condensate in imaginary time. The
total energies between states with and without a vortex are compared, yielding
critical rotation frequencies that depend on the anisotropy of the trap and
the number of atoms. Vortices displaced from the center of nonrotating traps
are found to have long lifetimes for sufficiently large numbers of atoms. The
relationship between vortex stability and bound core states is explored.
\end{abstract}
]

\narrowtext

The recent experimental achievement of Bose-Einstein condensation (BEC) in
trapped ultracold atomic vapors~\cite{Cornell,Ketterle,Hulet,Hau,Kleppner} has
provided a unique opportunity to investigate the superfluid properties of
weakly-interacting dilute Bose gases. Mean-field theories, which are usually
based on the Bogoliubov approximation~\cite{Bogoliubov,Fetter1} or its
finite-temperature extensions~\cite{Griffin}, yield an excellent description
of both the static and dynamic properties of the confined gases~\cite{Reviews}.
These theories also predict that a continuum Bose condensate with repulsive
interactions should be a superfluid, which can exhibit second sound, quantized
vortices, and persistent currents. While there exists some evidence for second
sound in trapped condensates~\cite{Ketterle2}, vortices in these systems have
never been observed despite considerable experimental effort~\cite{private}.
Numerous techniques for the generation of vortices have been suggested,
including stirring the condensate with a blue detuned
laser~\cite{McCann,Davies}, adiabatic population transfer via a Raman
transition into an angular momentum state~\cite{Marzlin,Dum}, spontaneous
vortex formation during evaporative cooling~\cite{Drummond,Choi}, and rotation
of anisotropic traps~\cite{Fetter2}.

Several studies of vortex stability recently have been carried
out~\cite{Fetter3,Rokhsar,Muller,Pu,Machida}. The free energy of a
singly-quantized vortex attains a local maximum when the vortex is
centered in a stationary trap~\cite{Fetter2}. In the presence of
dissipation, such a vortex would migrate to the edge of the trap
and eventually disappear~\cite{Rokhsar}. It has been
suggested~\cite{Fetter3,Rokhsar} that this mechanical instability may be
related to a bound state in the vortex core, corresponding to a
negative-energy `anomalous' dipole mode found numerically in the vortex state
at low densities~\cite{Dodd}. It is possible to stabilize singly-quantized
vortices by rotating the trap at an angular frequency $\Omega$. When
$\Omega$ is larger than the `metastability' frequency $\Omega_0$,
infinitesimal displacements of the vortex no longer decrease the system's
free energy, and the vortex becomes locally stable; above the critical
frequency $\Omega_c\ge\Omega_0$, the vortex state becomes the ground state
of the condensate~\cite{Fetter2}.

While rotation of the confining potential has been proposed as a method to both
nucleate and stabilize vortices in trapped Bose gases, the relevant critical
angular frequencies are not presently known. In the present work, numerical
results are obtained for Bose-condensed atoms confined in three-dimensional
rotating anisotropic traps at zero temperature. The critical frequency
$\Omega_c$ is found to increase with
the degree of anisotropy in the plane of rotation. In order to {\it nucleate}
vortices, however, the trapped gas must be rotated either more rapidly than
$\Omega_c$, or at temperatures above the BEC transition~\cite{Drummond,Choi}.
While vortices in nonrotating traps are found to be always unstable, for large
numbers of atoms their lifetimes can be very long compared with a trap period.

The trapped Bose condensate, comprised of $N_0$ repulsively-interacting Rb
atoms with mass $M=1.44\times 10^{-25}$~kg and scattering length
$a\approx 100a_0=5.29$~nm~\cite{Eite}, obeys the time-dependent
Gross-Pitaevskii (GP) equation in the rotating reference frame~\cite{GP}:

\begin{equation}
i\partial_{\tau}\psi({\bf r},\tau)
=\left[-{\case1/2}\vec{\nabla}^2+V_{\rm t}+V_{\rm H}-\Omega L_z\right]
\psi({\bf r},\tau),
\label{gp}
\end{equation}

\noindent where the trap potential is
$V_{\rm t}={\case1/2}\left(x^2+\alpha^2y^2+\beta^2z^2\right)$, the Hartree
term is $V_{\rm H}=4\pi\eta|\psi|^2$, and the condensate is rotated about the
$z$-axis at the trap center. The effects of gravity (along $\hat{z}$) are
presumed negligible. The constant rotation at frequency $\Omega$ induces
angular momentum per particle given by the expectation value of
$L_z=i\left(y\partial_x-x\partial_y\right)$. The trapping frequencies are
$(\omega_x,\omega_y,\omega_z)=\omega_x(1,\alpha,\beta)$ with
$\omega_x=2\pi\times 132$~rad/s, $\alpha\geq 1$, and
$\beta=\sqrt{8}$~\cite{Cornell}. Choosing the condensate to be
normalized to unity yields the scaling parameter $\eta=N_0a/d_x$. Note
that energy, length, and time are given throughout in scaled harmonic
oscillator units $\hbar\omega_x$,
$d_x=\sqrt{\hbar/M\omega_x}\approx 0.94~\mu$m, and
${\rm T}=2\pi/\omega_x\approx~7.6$ ms, respectively.

The ground-state of the GP equation is found within a discrete-variable
representation (DVR)~\cite{Feder1} by imaginary time propagation using an
adaptive stepsize Runge-Kutta integrator. A total of between $40\,000$ and
$130\,000$ DVR points of a Gauss-Hermite quadrature are used, and all
calculations are performed on a standard workstation. The stationary ground
state in the rotating frame is found by setting $\tilde{\tau}\equiv i\tau$ and
solving the diffusion equation:

\begin{equation}
\partial_{\tilde{\tau}}\psi({\bf r},\tilde{\tau})=-\left(H-\mu\right)
\psi({\bf r},\tilde{\tau}),
\label{gp2}
\end{equation}

\noindent where $H$ is the GP operator appearing on the right side of
Eq.~(\ref{gp}) and $\mu$ is the chemical potential. The condensate
wavefunction is assumed to be even under inversion of $z$, and is initially
taken to be the vortex-free Thomas-Fermi result, which is the time-independent
solution of Eq.~(\ref{gp2}), neglecting the kinetic energy operator and $L_z$.
A vortex is generated by imposing one quantum of circulation $\kappa$, a
$2\pi$-winding of the phase around the $z$-axis, on the condensate
wavefunction at $\tilde{\tau}=0$. At each imaginary timestep, the chemical
potential $\mu$ is readjusted in order to preserve the norm of the
wavefunction (i.e.\ the value of $N_0$). The propagation continues until the
right side of Eq.~(\ref{gp2}) is equal to a tolerance $\delta\leq 10^{-10}$
defining the error in the dimensionless chemical potential. Stationary
solutions are verified by subsequently integrating in real time; any
deviations from self-consistency would be made manifest by collective motion.

\begin{figure}[tb]
\centerline{\psfig{figure=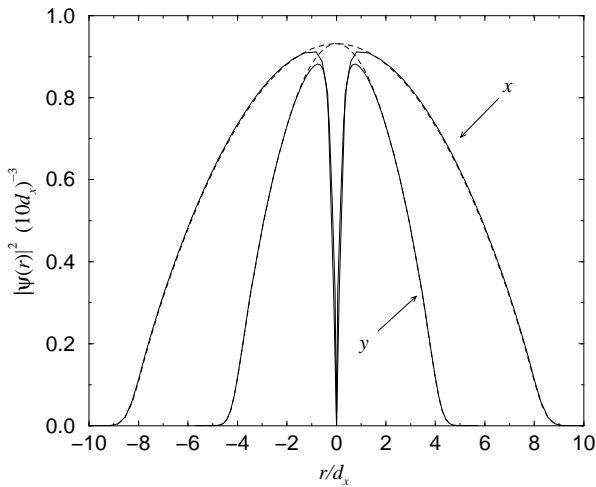,width=0.9\columnwidth,angle=270}}
\caption{The condensate density with a singly-quantized vortex at the origin
(solid lines), is shown projected along the $x$ and $y$-axes. The vortex-free
case is shown as dashed lines for comparison. Parameters for both cases are
$N_0=10^5$ and $\alpha=2$, yielding $\xi\approx 0.12d_x$.}
\label{vortex_fig}
\end{figure}

The solution to the GP equation for a vortex located at the center of a
nonrotating anisotropic trap containing $N_0=10^5$ atoms is shown in
Fig.~\ref{vortex_fig}. In general, the vortex core is found to become
decreasingly anisotropic as $N_0$ increases. Furthermore, the condensate
density preserves its overall vortex-free shape far from the origin except for
a slight overall bulge in order to preserve the norm. The structure of the
vortex indicates that the healing length $\xi$ is governed largely by the
local density and is only weakly dependent on trap geometry. In the
Thomas-Fermi (TF) approximation, which is valid for large $N_0$, the healing
length scales with the TF $\hat{x}$-axis radius
$R=(15\alpha\beta\eta)^{1/5}d_x$ as $\xi\sim(d_x/R)d_x$\cite{Fetter2}. Indeed,
the numerics clearly indicate that the mean vortex core radius (approximately
$d_x$ at low densities) shrinks very slowly as the TF limit is approached.

A superfluid subjected to a torque will remain purely irrotational until the
critical frequency $\Omega_c$ is reached, at which point it becomes globally
favorable for the system to contain a vortex with a single quantum of
circulation $\kappa$. In cylindrically-symmetric systems where the Hamiltonian
commutes with $L_z$, the circulation and angular momentum (with quantum $m$)
are identical; in the rotating frame, the free energies of the $m\neq 0$-states
are shifted by $m\Omega$, and $\Omega_c$ is simply the difference in energy
between the $m=1$ and $m=0$ states (divided by $\hbar$). In fully anisotropic
traps, however, even the $\kappa=0$ state is shifted, so the applied $\Omega$
in Eq.~(\ref{gp}) must be increased until the free energy curves cross. It is
straightforward to extend the TF estimate of $\Omega_c$~\cite{Sinha} to
include a small deviation from cylindrical symmetry~\cite{Fetter2}; neglecting
the shift of the vortex-free chemical potential (valid for $\alpha\sim 1$),
one obtains:

\begin{equation}
\Omega_c\approx{5\alpha\over 2}\left({d_x^2\over R^2}\right)
\ln\left({R\over\xi}\right)\omega_x.
\label{omegac}
\end{equation}

\begin{figure}[tb]
\centerline{\psfig{figure=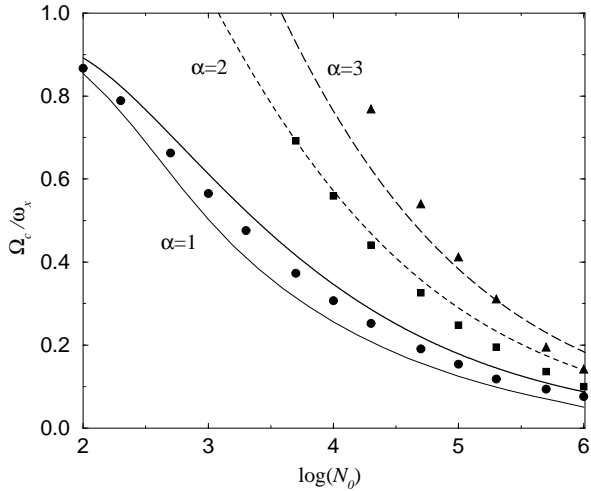,width=0.9\columnwidth,angle=270}}
\caption{The critical frequency for the global stabilization of a
singly-quantized vortex at the origin is given as a function of the number
of atoms in the condensate $N_0$ and in-plane anisotropy $\alpha$. Circles,
squares, and triangles show numerical results for $\alpha=1,2,3$; bold solid,
short dashed, and long dashed lines are the corresponding TF estimates,
respectively. The thin solid line represents the frequency of the anomalous
mode (shown as a positive value) in the vortex state for $\alpha=1$.}
\label{omegac_fig}
\end{figure}

Fig.~\ref{omegac_fig} shows the critical frequencies for the global stability
of a vortex with $\kappa=1$ at the trap center. For all geometries, the
critical frequency drops monotonically as $N_0$ is increased. For a given
number of atoms, the value of $\Omega_c$ increases with in-plane anisotropy,
similar to the behavior found for liquid helium in rotating elliptical
containers~\cite{Fetter4}. The energy of vortex formation must exceed that of
the irrotational velocity field, which is finite for a vortex-free condensate
in a rotating anisotropic trap. The TF result (\ref{omegac}) agrees well with
the numerical data in its regime of validity $\alpha\sim 1$, though it tends
to slightly overestimate the value of $\Omega_c$.

While $\Omega_c$ provides the criterion for the global stability of a vortex,
it does not necessarily indicate the critical frequency for vortex nucleation.
When initially vortex-free condensates are placed in anisotropic traps
rotating at a frequency $\Omega<\omega_x$, the velocity field of the stationary
solution is found to be irrotational even for $\Omega\gg\Omega_c$. In a
harmonic trap with smooth edges, it is not clear if there exists any suitable
locus for vortex formation. The vortices are most likely to originate at the
condensate surfaces normal to the axis of weak confinement, where the local
critical velocity is small~\cite{Kavoulakis} but the tangential superfluid
velocity in the laboratory frame is largest~\cite{Fetter4}. While these issues
are beyond the scope of the present issue of vortex stability, there is
evidence that multiple vortices appear at higher frequencies~\cite{Feder2}.
For smaller $N_0$, it would likely be easier to generate a vortex
experimentally by rotating the anisotropic trap before the condensate is
cooled below the BEC transition~\cite{Drummond,Choi}.

When $\alpha>1$, the angular momentum per particle $l_{\kappa}$ is a nontrivial
function of $N_0$, $\alpha$, and $\Omega$. In a nonrotating system with unit
vorticity, $l_{\kappa}$ increases with $N_0$. In the absence of a vortex,
$l_{\kappa}$ is finite for a given $\Omega$, and increases with $\alpha$;
the superfluid velocity ${\bf v}_s$ can be locally appreciable but
still remain irrotational $\nabla\times{\bf v}_s=0$. At the critical
frequency, the difference between $l_1$ and $l_0$ is always less than unity;
for the most extreme case considered here, a system with $N_0=10^6$ and
$\alpha=3$ rotating at $\Omega_c=0.14\omega_x$, one obtains $l_1=2.63\hbar$
and $l_0=1.77\hbar$. As $\alpha\to\infty$, the angular momentum approaches
that of a non-superfluid TF cloud $l_0\approx I_{\rm sb}\Omega$ with
`solid-body' moment of inertia $I_{\rm sb}={\case1/7}MR^2$.

An anisotropic harmonic oscillator potential becomes unconfining when it is
rotated at a frequency between the smallest and largest trapping frequencies.
Since $\Omega_c$ exceeds $\omega_x$ for sufficiently large $\alpha$, there
exists a critical minimum number of condensed atoms $N_c$ able to support a
vortex. The value of $N_c$ increases with $\alpha$ and is given by the
intercept of the $\Omega/\omega_x=1$ line in Fig.~\ref{omegac_fig}. In
cylindrically-symmetric systems $N_c=1$, since in the rotating frame the free
energies for all the $m$-states become degenerate at $\Omega=\omega_x$. In the
limit of extreme anisotropy $\alpha\to\infty$ vortices can never be stabilized.

It should be noted that states with vortices at the center of anisotropic
harmonic traps are found to be stationary solutions of Eq.~(\ref{gp2}) for 
all values of $N_0$ and $\Omega\geq 0$ considered; such configurations do not 
appear to decay in either real or imaginary time. A vortex at the center of a
nonrotating cylindrical trap increases the system's free energy, but is 
stationary because both the vorticity and angular momentum commute with the 
Hamiltonian; in principle, the angular momentum can be eliminated, and the
free energy reduced, only if this symmetry is broken by displacing the vortex
from the center. Since angular momentum is not conserved in anisotropic traps,
the apparent vortex stability is likely due to the free energy maximum at the
trap center~\cite{Fetter2}. In the absence of an external pinning mechanism,
any such configuration should be unstable against infinitesimal displacements.

In order to further explore the issue of vortex stability in nonrotating traps,
the initial condensate phase is wound by $2\pi$ a small distance
$x_0\approx 0.2d_x$ from the origin of a trap with $\alpha=1$. For all values
of $N_0\leq 10^6$, the condensate wavefunction rapidly (by
$\tilde{\tau}\sim$~T) converges to a metastable solution with a vortex, where
the fluctuations in $\mu$ become smaller than $\delta\approx 10^{-7}$ per
timestep $\Delta\tilde{\tau}\sim 10^{-3}$T. This wavefunction subsequently
decays to the true ground state, but both the real and imaginary time required
to do so is found to increase with $N_0$~\cite{Comment}. To an excellent
approximation, the total time diverges like $\tilde{\tau}\propto N_0^{2/5}$T;
for $N_0\mathrel{\lower1pt\hbox{$>\atop\raise2pt\hbox{$\sim$}$}}10^5$, the time
required $\sim 30$T becomes computationally inaccessible and the vortex state
becomes numerically indistinguishable from stationary. The numerics suggest
that while vortices in nonrotating traps are always unstable against
off-center displacements, they may be very long-lived.

The observed $x_0>0$ instability of the vortex state is likely due to the
existence of an `anomalous' collective mode $\omega_a$ at low
densities~\cite{Fetter2,Dodd,Rokhsar}. This dipole mode, which has positive
norm but negative energy (or vice versa), is associated with a zero angular
momentum bound state in the vortex core~\cite{Rokhsar}; its value corresponds
to the precession frequency of the vortex relative to the cloud~\cite{Fetter2}.
Previous numerical calculations~\cite{Dodd} found $|\omega_a|>0$ for all
$N_0\leq 10^4$. As the core radius shrinks with larger $N_0$, however, the
anomalous energy might be pushed to zero, yielding long-lived or even stable
vortices in the TF limit.

The low-lying excitation frequencies of a nonrotating condensate in the vortex
state are calculated using the Bogoliubov equations~\cite{Bogoliubov,Fetter1}.
For completely anisotropic geometries, however, the Bogoliubov operator is too
large to diagonalize explicitly. Calculations are therefore restricted to the
cylindrical case $\alpha=1$, where the vortex condensate is
$\psi\equiv\psi_1(\rho,z)e^{i\phi}$ and the quasiparticle amplitudes $u$ and
$v$ are labeled by $m$, the projection of the angular momentum operator $L_z$.
The Bogoliubov equations are then
\begin{equation}
\left(\matrix{\hat{O} & -V_{\rm H}\cr
V_{\rm H} & -\hat{O}'\cr}\right)
\left(\matrix{u_m\cr v_{2-m}\cr}\right)=\epsilon_m
\left(\matrix{u_m\cr v_{2-m}\cr}\right),
\label{bog}
\end{equation}

\noindent where 
$\hat{O}\equiv-{\case1/{2\rho}}{\case\partial/{\partial\rho}}\rho
{\case\partial/{\partial\rho}}-{\case1/2}{\case{\partial^2}/{\partial z^2}}
+{\case{m^2}/{2\rho^2}}+V_{\rm t}+2V_{\rm H}$,
$\hat{O}'\equiv\hat{O}+{\case{2(1-m)}/{\rho^2}}$, and $u_1=v_1=\psi_1$ when
$\epsilon_1=0$. In the $\hat{\rho}$-direction, the points of the DVR grid
correspond to those of Gauss-Laguerre quadrature, and the kinetic energy
matrix elements are obtained using the prescription of Baye and
Heenen~\cite{Baye}.

The anomalous mode $\omega_a$, which is labeled by $m=2$, is shown as a
function of $N_0$ in Fig.~\ref{omegac_fig}. The results indicate that
$0<|\omega_a|\leq\Omega_c$ for all $N_0\leq 10^6$ considered. Our calculations
suggest that for large numbers of atoms, $\omega_a$ coincides with the
metastability rotation frequency $\Omega_0$ discussed above; the numerical
value of $\omega_a$ is consistent with the TF result
$\Omega_0={\case 3/5}\Omega_c$~\cite{Fetter2}. Indeed, in the frame of a
condensate rotating at $\Omega=\omega_a$, the frequency of the vortex
oscillation would be Doppler shifted to zero. Alternatively, it can be shown
in both the weakly-interacting and TF limits that $\Omega_0$ is also the
frequency at which the chemical potentials $\mu$ for the vortex and
vortex-free states become equal; in the TF limit, $\omega_a$ vanishes when the
vacua (or energy zero) for quasiparticle excitations for both states coincide.

Since the anomalous mode corresponds to the precession of the vortex about the
trap origin, one may make a crude estimate of the vortex lifetime $\tau$. In
the presence of dissipation, the vortex will spiral out of the condensate
after a few orbit periods $\omega_a^{-1}$. Assuming that
$\omega_a={\case 5/3}\Omega_c$, then with Eq.~(\ref{omegac}) one obtains
$\tau\sim N_0^{2/5}$T in the TF limit neglecting logarithmic factors. This
result is consistent with the imaginary time $\tilde{\tau}$ required to yield
the vortex-free ground state in the fully three-dimensional numerical
calculations discussed above. Similar decay times have been obtained for 
solitons and vortices in the presence of a small noncondensate
component~\cite{Shlyapnikov}.

In summary, we have obtained numerically the critical frequencies $\Omega_c$
for the stabilization of a vortex at the center of a rotating
anisotropically-trapped Bose condensate. Since $\Omega_c$ increases with the
in-plane anisotropy $\alpha=\omega_y/\omega_x$ and the condensate becomes
unconfined for $\Omega>\omega_x$, there is a minimum number of atoms able to
support a vortex state. Vortices in nonrotating traps are found to be unstable
against small off-center displacements, but their decay time diverges with the
total number of atoms.

\begin{acknowledgments}
The authors are grateful for many stimulating discussions with M.~Brewczyk,
J.~Denschlag, M.~Edwards, A.~L.~Fetter, D.~Guery-Odelin, D.~A.~W.~Hutchinson,
M.~Matthews, W.~D.~Phillips, W.~Reinhardt, and C. J. Williams. This work was
supported by the U.S.\ Office of Naval Research.
\end{acknowledgments}


\begin{references}
\bibitem{Comment}
Since the GP equation conserves energy and contains no explicit damping terms,
one might expect the displaced vortex to precess about the origin indefinitely.
In fact, the vortex gradually spirals toward the low-density surface in real
time, inducing collective motion of the condensate in the process.
\end{references}
\end{document}